\documentclass[journal=jctc,manuscript=article]{achemso}
\usepackage[version=3]{mhchem} 
\usepackage{graphicx, color, float, array}
\usepackage{amsfonts}
\usepackage{xr}
\usepackage{url}
\makeatletter
\newcommand*{\addFileDependency}[1]{
  \typeout{(#1)}
  \@addtofilelist{#1}
  \IfFileExists{#1}{}{\typeout{No file #1.}}
}
\makeatother

\newcommand*{\myexternaldocument}[1]{%
    \externaldocument{#1}%
    \addFileDependency{#1.tex}%
    \addFileDependency{#1.aux}%
}
\myexternaldocument{supp_info}

\newcommand*{\citen}[1]{%
  \begingroup
    \romannumeral-`\x 
    \setcitestyle{numbers}%
    \cite{#1}%
  \endgroup   
}

\author{Dhiman Ray}
\affiliation[]
{Atomistic Simulations, Italian Institute of Technology, Genoa, Via Enrico Melen 83, GE 16153, Italy }
\author{Enrico Trizio}
\affiliation[]
{Atomistic Simulations, Italian Institute of Technology, Genoa, Via Enrico Melen 83, GE 16153, Italy }
\alsoaffiliation[]
{Department of Materials
Science, Universit{\`a} di Milano-Bicocca, Milano 20126, Italy}
\author{Michele Parrinello}
\email{michele.parrinello@iit.it}
\affiliation[]
{Atomistic Simulations, Italian Institute of Technology, Genoa, Via Enrico Melen 83, GE 16153, Italy }

\title[]
{Deep Learning Collective Variables from Transition Path Ensemble}
\abbreviations{IR,NMR,UV}

\begin{document}


\newpage
\begin{abstract}
The study of the rare transitions that take place between long lived metastable states is a major challenge in molecular dynamics simulations. Many of the methods suggested to address this problem rely on the identification of the slow modes of the system which are referred to as collective variables. Recently machine learning methods have been used to learn the collective variables as functions of a large number of physical descriptors. 
 Among  many such methods   Deep Targeted Discriminant Analysis has proven to be useful. This collective variable is built from data harvested in short unbiased simulation in the two basins. Here we enrich the set of data on which the Deep Targeted Discriminant Analysis collective variable is built by adding data coming from the transition path ensemble.  These are  collected from a number of reactive trajectories obtained using the On-the-fly Probability Enhanced Sampling Flooding method.  The collective variables thus trained,  lead to a more accurate sampling and faster convergence.  The performance of these new collective variables is tested on a number of representative examples. 

\end{abstract}

\section{Introduction}
Enhanced sampling methods have gained significant popularity in molecular dynamics (MD) simulations since they enable studying rare events taking place on a computationally unaffordable time scale\cite{Henin2022enhanced,yang2019enhanced,kamenik2022enhanced}. The fundamental principle behind most enhanced sampling methods is the application of an external bias to drive rare transitions, whose timescales would otherwise remain outside the scope of standard simulations \cite{yang2019enhanced}. Apart from a few exceptions \cite{mitsutake2001generalized,hamelberg2004accelerated}, such bias potential is usually defined as a function of a small number of collective variables (CVs) $\mathbf{s} = \mathbf{s}(\mathbf{R})$ which are functions of the atomic coordinates $\mathbf{R}$ and are meant to encode the slow modes of the system. 

Since the introduction of the CV-based biasing scheme by Torrie and Valleau \cite{torrie1977nonphysical}, a large number of different enhanced sampling algorithms have been developed using this principle \cite{schulten1983steered,carter1989constrained,darve2002calculating,wolf2018targeted,laio2002escaping,barducci2008well,valsson2014variational,invernizzi2020rethinking,invernizzi2022exploration}. Our group has been active in this area and has introduced  first metadynamics \cite{laio2002escaping} and later its improved variant the On-the-fly Probability Enhanced Sampling (OPES) \cite{invernizzi2020rethinking}.  In both cases, the bias potential $V(\mathbf{s})$ is adaptively learned during the simulation by estimating the probability distribution in the CV space. 

Since the efficacy of these approaches depends crucially on the  CV a large body of work has been devoted to designing effective CVs that could drive transitions across energy barriers and allow the underlying free energy landscape to be calculated in an efficient manner \cite{branduardi2007b,leines2012path,piaggi2017entropy,ravindra2020automatic,tiwary2016spectral,ribeiro2018reweighted,mendels2018collective,grifoni2019microscopic,zhang2019improving,sultan2017tica,sultan2018automated,mccarty2017variational,bonati2020data,bonati2021deep,trizio2021enhanced}.

In the traditional approach, the CVs are taken as functions of few carefully selected degrees of freedom, like for instance interatomic distances, torsion angles, and coordination numbers. This approach offers the advantage of a transparent physical interpretation, but it can fail to capture  the complex behavior of many molecular systems.

In recent years, different data-based approaches have been applied to the CVs construction. One class of such methods aims at identifying the slowest modes of the systems.  We quote here the approach based on the principal component analysis\cite{amadei1993essential} or on the more complex  Time-lagged Independent Component Analysis TICA \cite{molgedey1994separation,perez2013identification,schwantes2013improvements}.  The power of these linear methods has been boosted  by the  use of  Neural Networks (NNs) that take advantage of NNs ability  to approximate non linear functions of many variables. This has led to the development of efficient CVs such as those that are built using the  Reweighted Autoencoded Variational Bias (RAVE) \cite{ribeiro2018reweighted} or the Deep-TICA \cite{bonati2021deep}. 

A different principle in CV design has been to use  classification methods to build coordinates that can distinguish between the different metastable states of interest (e.g. the folded, unfolded, and (or) the misfolded states of a protein).  In this case, the data are generated by performing  unbiased simulations in the different  metastable states.   A frequently used linear method that is based on a classification approach is the Harmonic Linear Discriminant Analysis (HLDA) \cite{mendels2018collective}  which has found applications to  chemical reactions \cite{piccini2018metadynamics,rizzi2019blind}. As in some of the methods described earlier NNs have later been utilized to improve upon these CVs. Examples are  Deep-LDA (Deep Linear Discriminant Analysis) \cite{bonati2020data} and Deep-TDA (Deep Targeted Discriminant Analysis) \cite{trizio2021enhanced}. The application of these methods has been encouraging in a wide range of problems including the folding of small proteins \cite{ray2022rare}, ligand-receptor binding \cite{trizio2021enhanced,rizzi2021role,Ansari2022WaterTrypsin}, and phase transition in solid materials \cite{karmakar2021collective}. 


In this work, we aim at improving the efficiency of the Deep-TDA approach which combines the discrimination criterion with the requirement that the data, when projected along the CV, are distributed according to a preassigned target distribution. 
Our work is based on the recognition that a good CV should be able not only to distinguish between initial and final state but also to pass through the lowest free energy transition pathways. If these conditions are satisfied the bias will encourage the system to pass through the physical transition state and not through a higher free energy pathway.
This will lead to faster convergence and to a more accurate estimate of the free energy landscape. 


Guided by these considerations, we generalize the Deep-TDA classification method by adding a new set of data that comes from the transition path ensemble (TPE).
The TPE data are considered as a new class and are generated by using the recently developed OPES-Flooding\cite{ray2022rare} approach that has proven to be efficient in generating unbiased transition paths. 

We test our approach on the M{\"u}ller potential, the folding and unfolding of chignolin, and the association and dissociation of a host-guest complex. This new approach, which we call Transition Path Informed Deep-TDA (TPI-Deep-TDA), leads to significant improvements in the convergence speed and the accuracy and precision of the computed free energy differences between relevant states.  

\section{Theory}
    \subsection{Deep Targeted Discriminant Analysis (Deep-TDA)}
        The Deep Targeted Discriminant Analysis\cite{trizio2021enhanced} (Deep-TDA) method, which was developed for the data-driven design of CVs starting from the description of the metastable states of a system. The Deep-TDA CVs are built as the output of a feed-forward Neural Network (NN) which is optimized following a discrimination criterion and takes as inputs a large set of physical descriptors collected from short unbiased simulations that explore only metastable basins. Such descriptors should be invariant with respect to the symmetries of the system. Typical examples of descriptors used in the practice are interatomic distances, angles, or coordination numbers.
        
        Given a system with $N_m$ metastable states that can be characterized by a set of $N_d$ descriptors \textbf{d}, the NN is optimized to map the multi-dimensional space of descriptors \textbf{d} into a $N_s$ dimensional CV \textbf{s}. 
        The NN is trained so that the data from each metastable state, when projected along the CV, are distributed according to a preassigned target in which the different states are well-defined. In the practice, this target is taken as a sum of $N_m$ of Gaussians, one for each state that we want to classify.
        
        The rationale for choosing this targeted approach is as follows. Since the data are well separated in the physical configuration space, a straightforward application of a discrimination criterion will cause the metastable state distributions to be sharply peaked and distant in the CV space.
        The resulting  CVs would then have a very strong dependence on the atomic coordinates $R$ which is not suitable in a biasing context. Contrarily, the imposition of our target allows the CV to better reflect the physical distribution of the data.
        
        Since the target is chosen to be a linear combination of $N_m$ Gaussians of preassigned positions and widths, the loss function is given by:
        
        \begin{equation}
                L = \alpha \sum_{k}^{N_s}\sum_{l}^{N_d} (\mu_{k,l} -\overline{\mu}_{k,l})^2 +\beta \sum_{k}^{N_s}\sum_{l}^{N_d} (\sigma_{k,l} -\overline{\sigma}_{k,l})^2
                \label{eq:loss_multi}
            \end{equation}

        where the first term enforces that the average positions $\mu_{k,l}$ of the data in metastable state $k$ for the $l$ component of $\mathbf{s}$ are close to the center of the target Gaussians $\overline{\mu}_{k,l}$ and similarly the second term makes the spreads in the different metastable states $\sigma_{k,l}$ close to the target ones $\overline{\sigma}_{k,l}$.
        The hyperparameters $\alpha$ and $\beta$ regulate the relative weights of the center and sigma-related terms and a judicious choice of their values improves the optimization procedure. 
        A more detailed description of the optimization procedure can be found in Ref.\cite{trizio2021enhanced}.
        
    \subsection{On-the-fly Probability Enhanced Sampling (OPES)}
    To accelerate  sampling we employ the On-the-fly Probability Enhanced Sampling (OPES) method developed by Invernizzi and Parrinello \cite{invernizzi2020rethinking}. 
    In OPES one preassigns a target distribution for  the collective variables $p^{\text{tg}}(\mathbf{s})$.  In principle  one can  choose freely $p^{\text{tg}}(\mathbf{s})$, however,  most of the time one chooses as target the well tempered distribution that is related to the unbiased marginal probability distribution $P(\mathbf{s})$ by $p^{\text{tg}}(\mathbf{s}) \propto [P(\mathbf{s})]^{1/\gamma}$, $\gamma >1$ being the bias factor. 
    With this choice of the target function, the bias at the $n$-th iteration is written as 
    \begin{equation}
        V_n(\mathbf{s})=(1-1/\gamma)\frac{1}{\beta} \ln \left( \frac{P_n(\mathbf{s})}{Z_n}+\epsilon \right)\, ,
    \end{equation}
    where $P_n(\mathbf{s})$ is the estimated $P(\mathbf{s})$ at iteration $n$. 
    Gaussian kernels are used to reconstruct $P_n(\mathbf{s})$. $Z$ is a normalization factor and $\epsilon$ is a regularization term that is included to ensure the numerical stability of the algorithm.  OPES has proven to be rather efficient in the study of a variety of rare events \cite{invernizzi2022exploration,Ansari2022WaterTrypsin,rizzi2021role,raucci2022discover,ray2022rare,karmakar2021collective}.

    In the present context, in which we are interested in collecting transition path data, OPES recommends itself also because of its variant called OPES-Flooding which allows computing reaction rates and 
    harvest transition paths\cite{ray2022rare}.

    OPES-Flooding, similar to infrequent metadynamics, avoids depositing bias in the transition region so as to recover reaction rates. 
    In OPES-Flooding this objective is obtained by imposing that no bias is deposited for $s>s_{exc}$. 
    A careful choice of the so-called excluded region parameter $s_\mathrm{exc}$ ensures that no bias is introduced in the transition state region.  At the same time, the bias still introduced for $s < s_\mathrm{exc}$ will accelerate the probability  of observing a transition. We note that the choice of an appropriate $s_{exc}$ becomes natural when using a Deep-TDA CV as the metastable state distributions are localized in predefined regions.
    The reaction paths generated using OPES-Flooding can be used to obtain unbiased data on the transition path ensemble. Alongside this, the transition rates can also be easily computed.

    
    \subsection{Transition Path Informed Deep Targeted Discriminant Analysis (TPI-Deep-TDA) }
    It is commonly known that NNs are extremely powerful when used for interpolating from data but can lead to poor results when used for extrapolation. 
    Indeed, in Deep-TDA and other discriminant-based CVs, one assumes that a model which is trained to discriminate between the metastable states only will also provide a meaningful description of the transition state region. This is in general not a bad assumption since  the extrapolation takes into account the fact that the CV has to join smoothly the metastable states regions. However, sometimes the performance of such a CV can be far from optimal. 

\begin{figure}[b!]
        \centering
        \includegraphics[width=\textwidth]{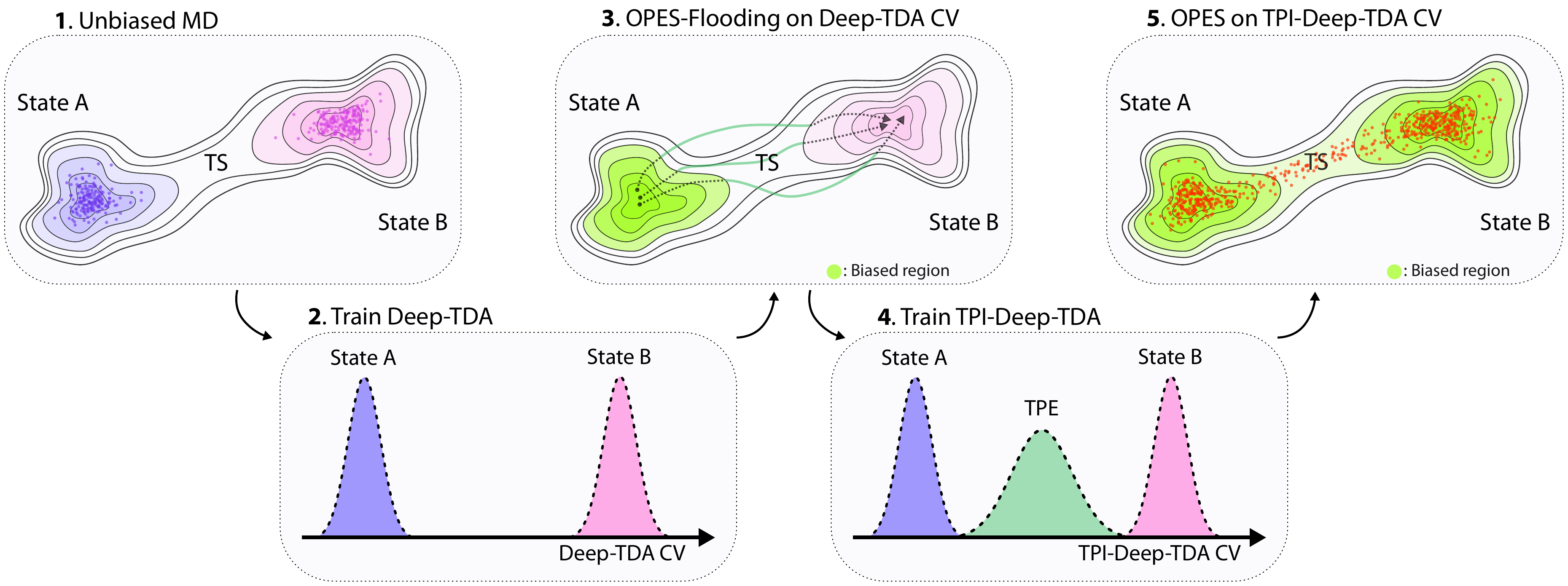}
        \caption{A schematic representation of the TPI-Deep-TDA CV construction for a two-state system. \textbf{1.} A set of physical descriptors \textbf{d} is collected by unbiased MD runs in the metastable basins of the system \,\textbf{2.} A Deep-TDA CV is trained such that the data from states A and B are distributed according to two Gaussians in the CV space \,\textbf{3.} Unbiased reactive trajectories are sampled using a set of OPES-Flooding\cite{ray2022rare} simulations along the Deep-TDA CV. The bias (green shade) is deposited only in one of the basins and excluded from the TS region. Only the sections of each reactive path that fall outside the metastable basins are taken as part of the TS-region dataset (marked in green) \,\textbf{4.} A second TPI-Deep-TDA CV is trained to fit the TPE data distribution to a third wider Gaussian (painted in green) between the metastable states A and B \,\textbf{5.} The TPI-Deep-TDA CV is biased in OPES\cite{invernizzi2020rethinking} to drive transitions between A and B applying bias (green shade) along the TS path.
        }
        \label{fig:schematic}
    \end{figure}

    To improve the quality of the Deep-TDA CV we propose to incorporate information from TPE obtained from reactive trajectories. The CV design protocol is depicted below for an example two-state system (Fig.\ref{fig:schematic}).


    \begin{itemize}
        \item  \textbf{Step 1:} We collect data on a set of descriptors by running unbiased simulations in the metastable basins of the system (Fig.\ref{fig:schematic} panel 1).
        
        \item \textbf{Step 2:} The  data collected in the metastable states are used  to train a standard Deep-TDA CV as discussed above \cite{trizio2021enhanced} (Fig.\ref{fig:schematic} panel 2).
        
        \item \textbf{Step 3:} Using the Deep-TDA CV thus generated we perform a set of OPES-Flooding\cite{ray2022rare} simulations and harvest several reactive trajectories. We select from the reactive trajectories only those configurations that lie  outside the metastable basins (Fig.\ref{fig:schematic} panel 3).
        
        \item \textbf{Step 4:} The new configurations thus obtained are added to the initial dataset and we train a new CV that we call Transition Path Informed Deep-TDA (TPI-Deep-TDA). 
        To do so, we modify the target distribution used for the Deep-TDA CV in Step 2 by adding a third wider Gaussian, placed between the ones related to the metastable states, and optimizing the NN to fit the TPE data distribution in the CV space to such Gaussian (Fig.\ref{fig:schematic} panel 4).
        
        \item \textbf{Step 5:} The TPI-Deep-TDA CV is finally used to perform OPES simulations to calculate the free energy landscape (Fig.\ref{fig:schematic} panel 5).
    \end{itemize} 
    

    The rationale behind the different choice of Gaussians in the target  approach is to reflect the structure of the physical data. Let us for simplicity consider only a two state case A and B.  In this case, the data can be divided  into three groups: those that belong to basin A, those that belong to basin B, and those that belong to the transition paths ensemble (TPE). The data in A and B  are localized at different positions in the high-dimensional descriptors space, therefore a mapping into separately localized Gaussians is a natural one. In contrast, the TPE data are spread across the region in between the metastable states.  To best mimic this structure, while still sticking to a Gaussian representation, we introduce a third Gaussian centered between the A and the B Gaussians. The corresponding width $\sigma_{TPE}$  is larger than those of A and B, bridging the intermediate region, and has a negligible overlap with either A and B Gaussians. 


    At this stage, we point out that the transition path ensemble used for training our proposed CV can, in principle, be sampled using various alternative schemes including transition path sampling \cite{dellago1998transition}, aimless shooting \cite{mullen2015easy}, transition interface sampling \cite{van2003novel}, metadynamics of paths \cite{mandelli2020metadynamics}, etc. It should also be possible to collect configurations specific to the TS region by applying the adaptive bias enhanced sampling with customized target distribution proposed by Debnath et al. \cite{debnath2019enhanced}. Nonetheless, we choose to use the OPES-Flooding algorithm as it does not require any pre-existing knowledge on the location of the TS, a piece of information that is not readily available. In addition, we also get the advantage of conveniently recovering the kinetics of the process.
    \section{Summary of Computational Methods}
    We tested our protocol   on the two-dimensional  M{\"u}ller-Brown potential, the folding and unfolding of  chignolin, and the binding of a ligand to an octa-acid host.
    A detailed description of the computational details can be found in the SI. Here we report only an abridged set of information.
    Langevin dynamics simulations for the M{\"u}ller-Brown potential were performed using PLUMED 2.9\cite{plumed2019promoting,tribello2014plumed}, and classical MD simulations for the chignolin and the host-guest system were performed using the Cuda-enabled version of GROMACS v2021.5 \cite{abraham2015gromacs} patched with the PLUMED v2.9 \cite{plumed2019promoting,tribello2014plumed}. The CHARMM22$^*$ force field \cite{piana2011robust} and the Generalized AMBER Force Field (GAFF) \cite{wang2004development} were used to model the chignolin and the host-guest system respectively. All neural networks were trained using the \texttt{mlcvs} package (\url{https://github.com/luigibonati/mlcvs}) implemented in PyTorch. Biasing the NN CVs was accomplished using the PLUMED-PyTorch interface \cite{bonati2020data}.

    \section{Results and Discussions}
    \subsection{M{\"u}ller-Brown Potential}
    The two-dimensional M{\"u}ller-Brown potential is often used to test the efficiency of enhanced sampling methods.
    As we shall see below, it is a system in which a standard Deep-TDA CV performs rather well. Still, the extension of the Deep-TDA approach to include  transition path data is able to further improve the CV performance and speed up convergence. 
    
\begin{figure*}[t!]
        \begin{minipage}{0.49\linewidth}
            \includegraphics[width=\linewidth]{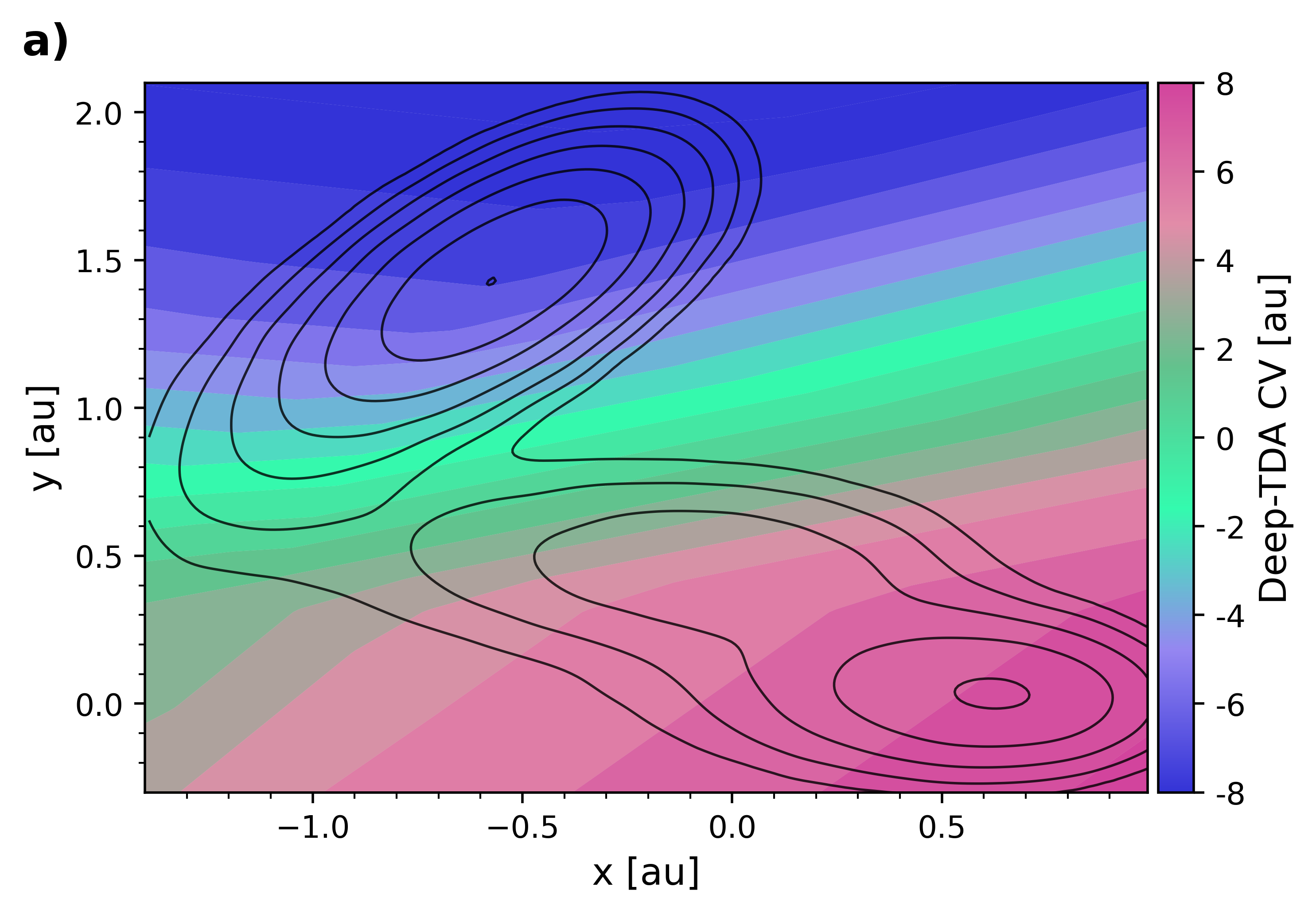}
        \end{minipage}
        \hspace*{\fill}
        \begin{minipage}{0.49\linewidth}
        \includegraphics[width=\linewidth]{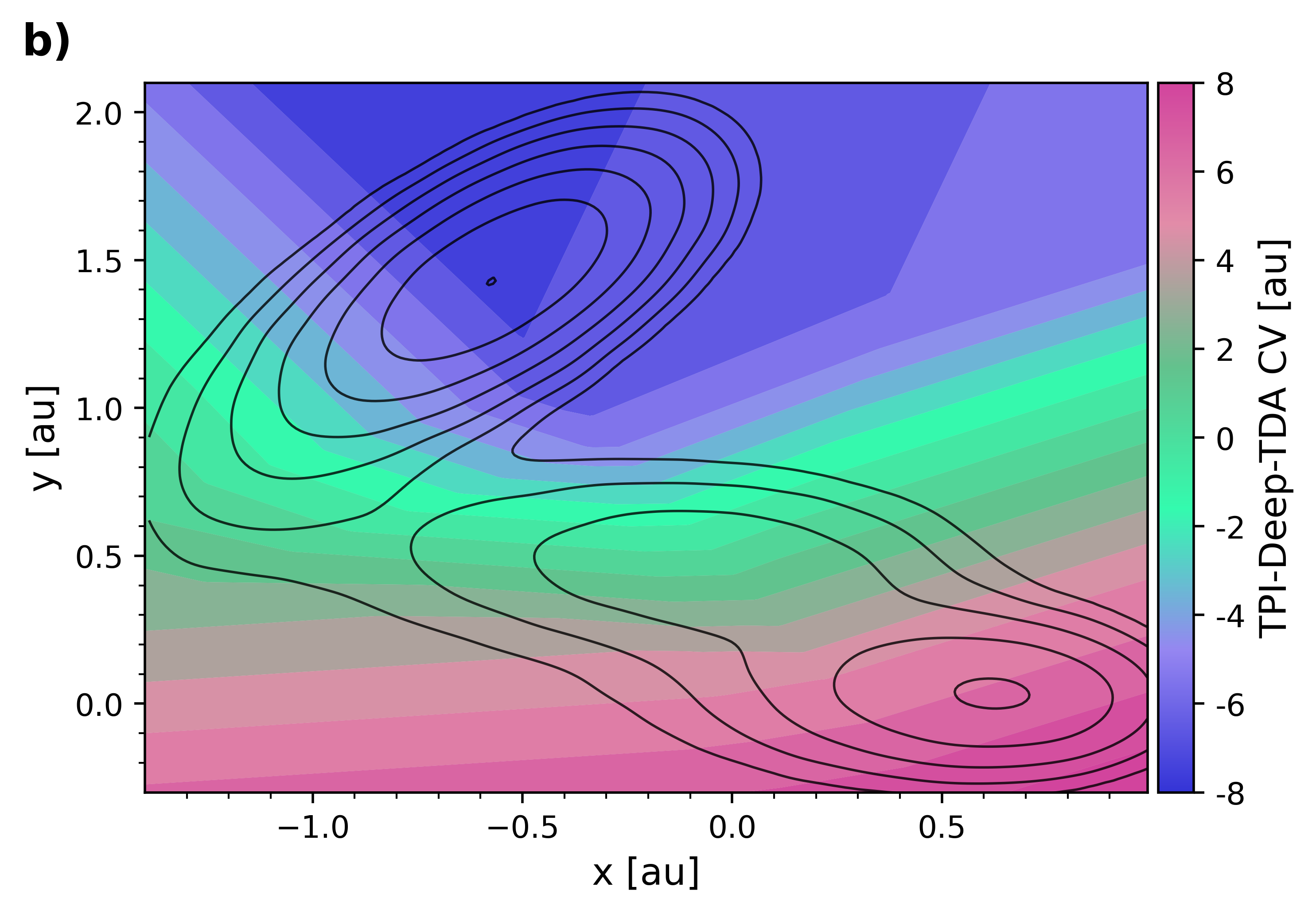}
        \end{minipage}
        \caption{The contour plots of   standard Deep-TDA  \textbf{(a)}and  TPI-Deep-TDA  \textbf{(b)}CVs for the M{\"u}ller-Brown potential. These CVs describe poorly the regions that are far from the training data. This confirms the inability of NN to extrapolate to data-poor regions. However, since these regions are not physically interesting, this is of little practical consequence.}
        \label{fig:mueller_cv_isoline}
    \end{figure*}

\begin{figure*}[t!]
        \begin{minipage}{0.49\linewidth}
            \includegraphics[width=\linewidth]{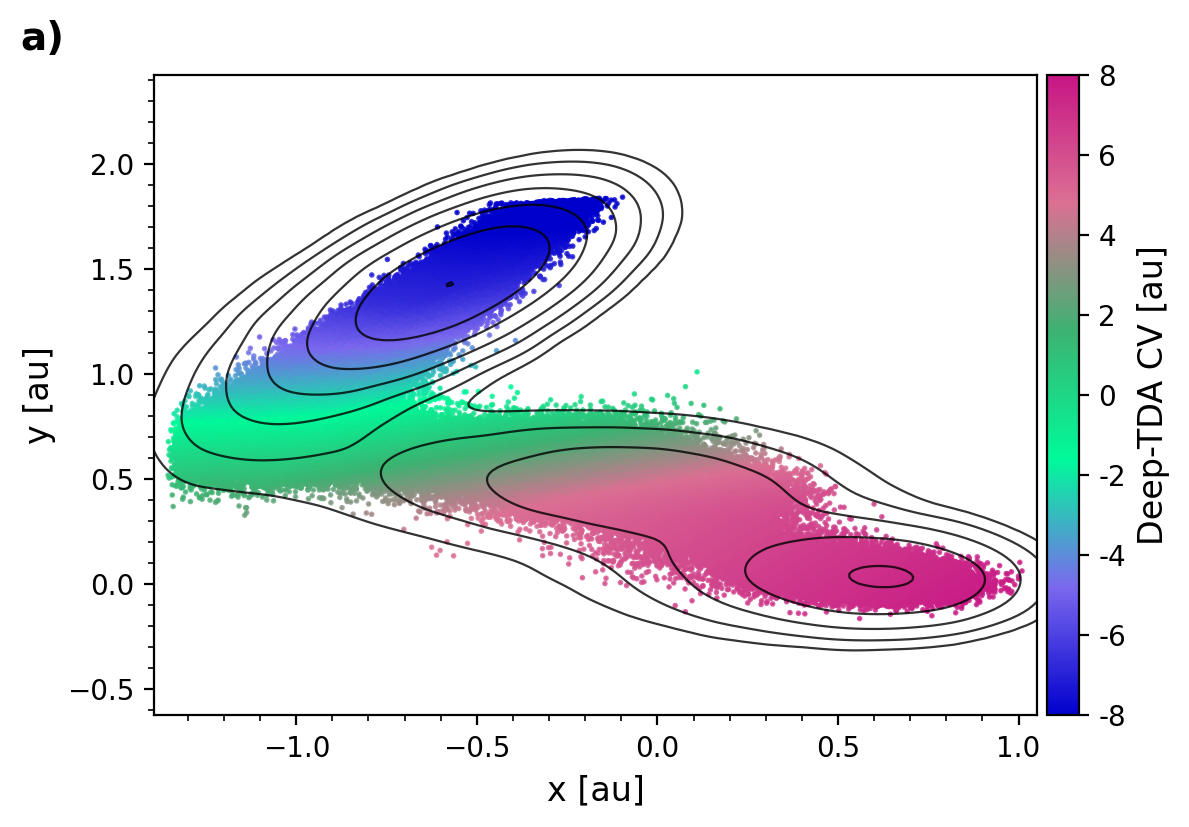}
        \end{minipage}
        \hspace*{\fill}
        \begin{minipage}{0.49\linewidth}
            \includegraphics[width=\linewidth]{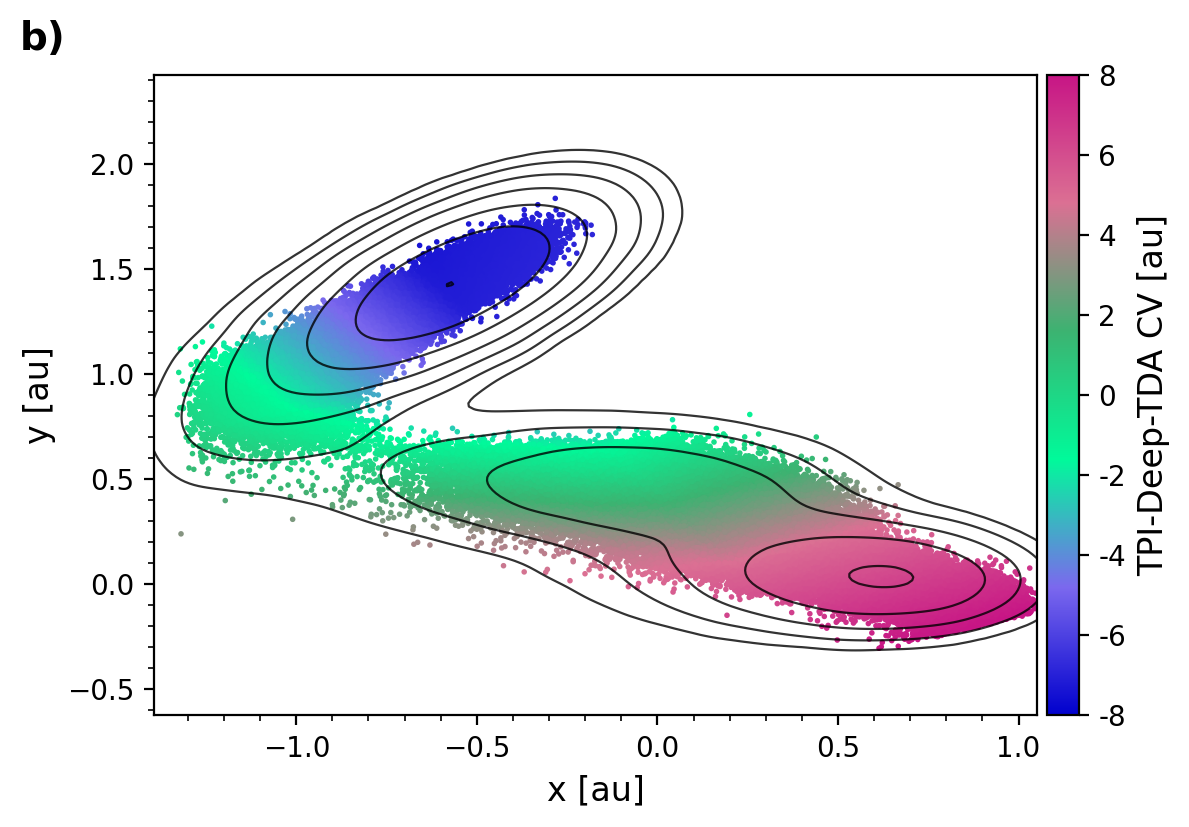}
        \end{minipage}
        \begin{minipage}{\linewidth}
            \centering
            \includegraphics[width=0.5\linewidth]{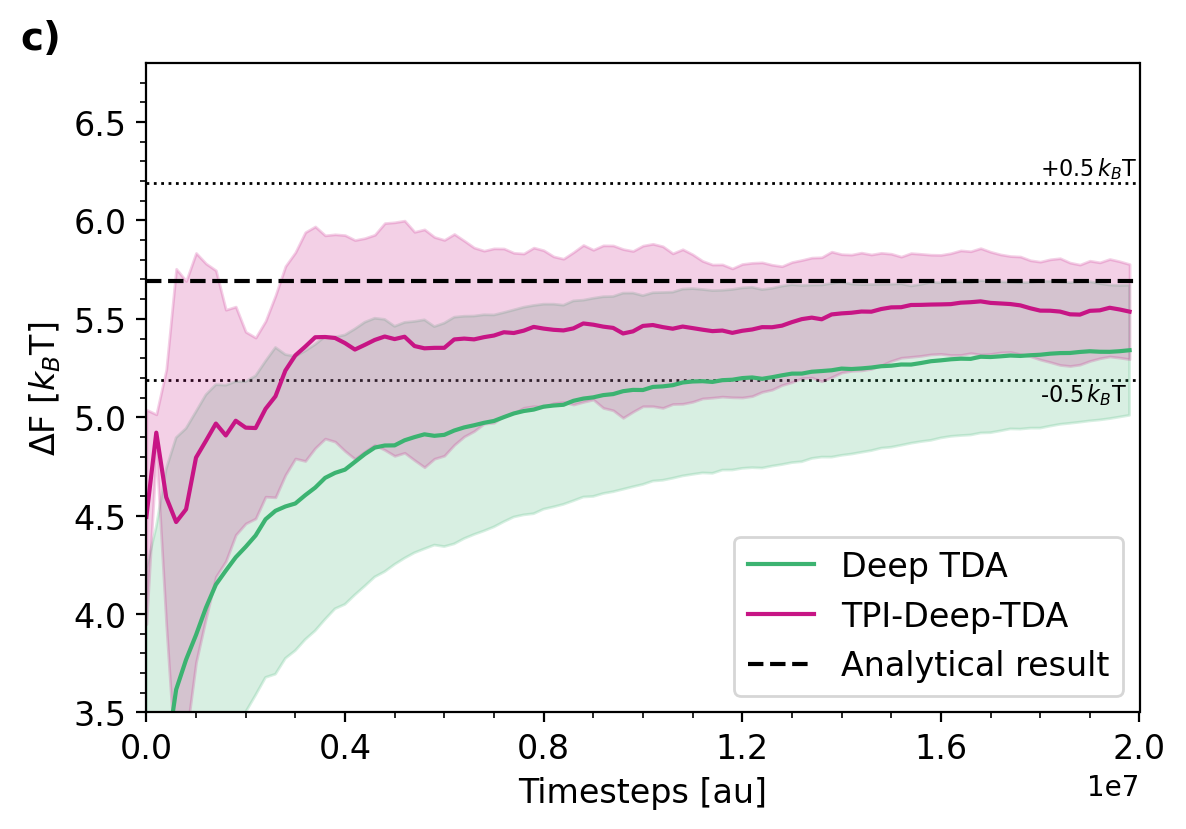}
        \end{minipage}
         \caption{Scatter plot of the points visited performing OPES simulations using \textbf{(a)} Deep-TDA and (\textbf{b}) TPI-Deep-TDA CVs on the M{\"u}ller-Brown potential, whose isolines are given in black. The colormap indicates the value of the corresponding CV. \textbf{(c)} Convergence of the free energy difference $\Delta F$  between the basins with simulation time. The solid line and the shaded region report respectively the average and standard deviation computed from three independent trajectories. The reference value of 5.69 k$_B$T was obtained by numerical integration\cite{invernizzi2022exploration} and the dotted lines mark the $\pm$0.5k$_B$T range around that value.}
         \label{fig:mueller_results}
    \end{figure*}

    In order to understand this different behavior we compare in Fig.\ref{fig:mueller_cv_isoline} Deep-TDA and TPI-Deep TDA CVs. In the case of Deep-TDA the CV can distinguish well between the metastable states 
    but it does not follow precisely the gradient of the underlying energy landscape. (Fig.\ref{fig:mueller_cv_isoline}, panel a).  On the other hand, the 
    TPI-Deep-TDA CV isolines follow the free energy gradient more closely (Fig.\ref{fig:mueller_cv_isoline}, panel b).
    As a consequence in the first case, the system is pushed by the bias to explore a larger than necessary portion of the transition state region and less relevant regions. Secondly, the nature of the transition state region is more closely encoded in the CV. This  facilitates the transitions between the metastable states and the points sampled remain closer to the minimum free energy path (Fig.\ref{fig:mueller_results}, panel b).(Fig.\ref{fig:mueller_results}, panel a). Thus convergence is much faster using TPI-Deep-TDA (Fig.\ref{fig:mueller_results}, panel c).
     
\subsection{Chignolin}
To study the folding and unfolding of chignolin, we trained two TPI-Deep-TDA CVs using two different descriptors sets. The first set is composed of all the pairwise contacts between the C$_\alpha$ atoms. The second set includes a curated set of interatomic distances suggested in Bonati et al.\cite{bonati2021deep}. The results from the latter case, are included in the SI.
For each set we perform three independent 1$\mu$s long OPES simulations.

\begin{figure*}[t!]
    \begin{minipage}{0.49\linewidth}
        \includegraphics[width=\linewidth]{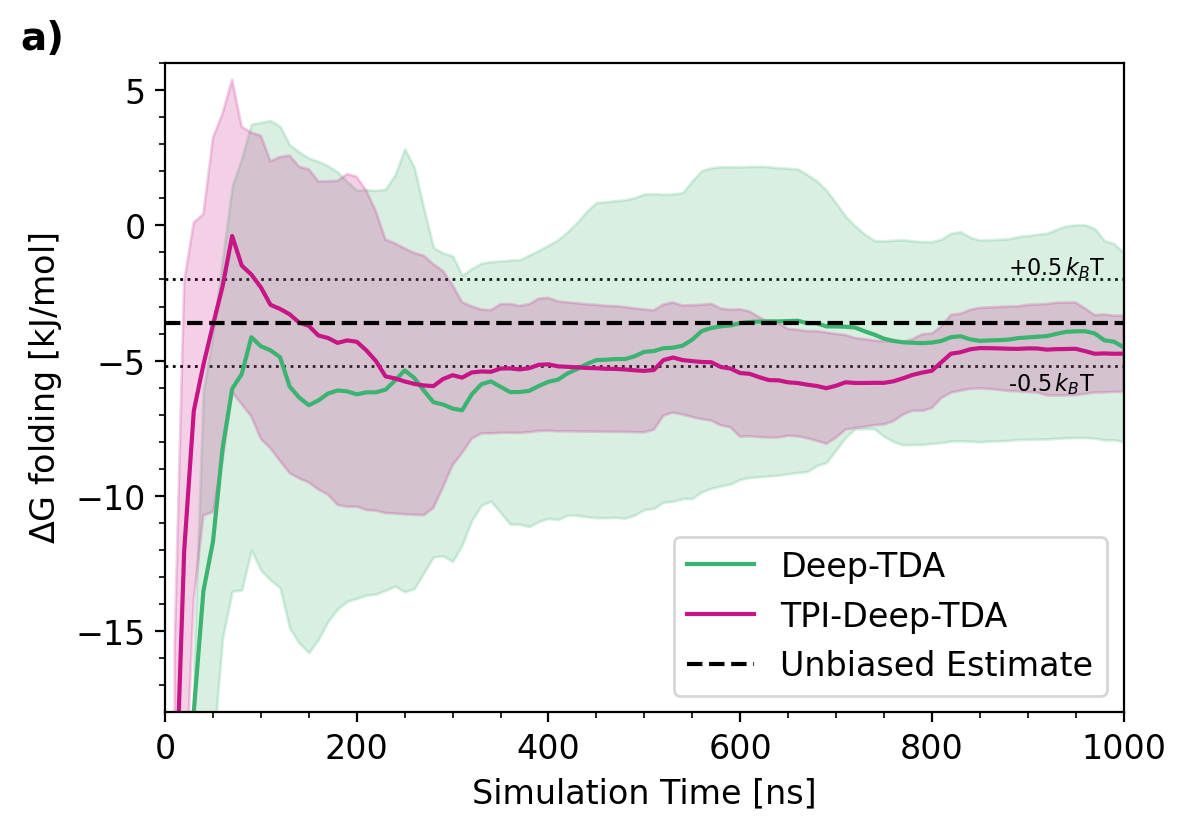}
    \end{minipage}
    \hspace*{\fill}
    \begin{minipage}{0.49\linewidth}
    \includegraphics[width=\linewidth]{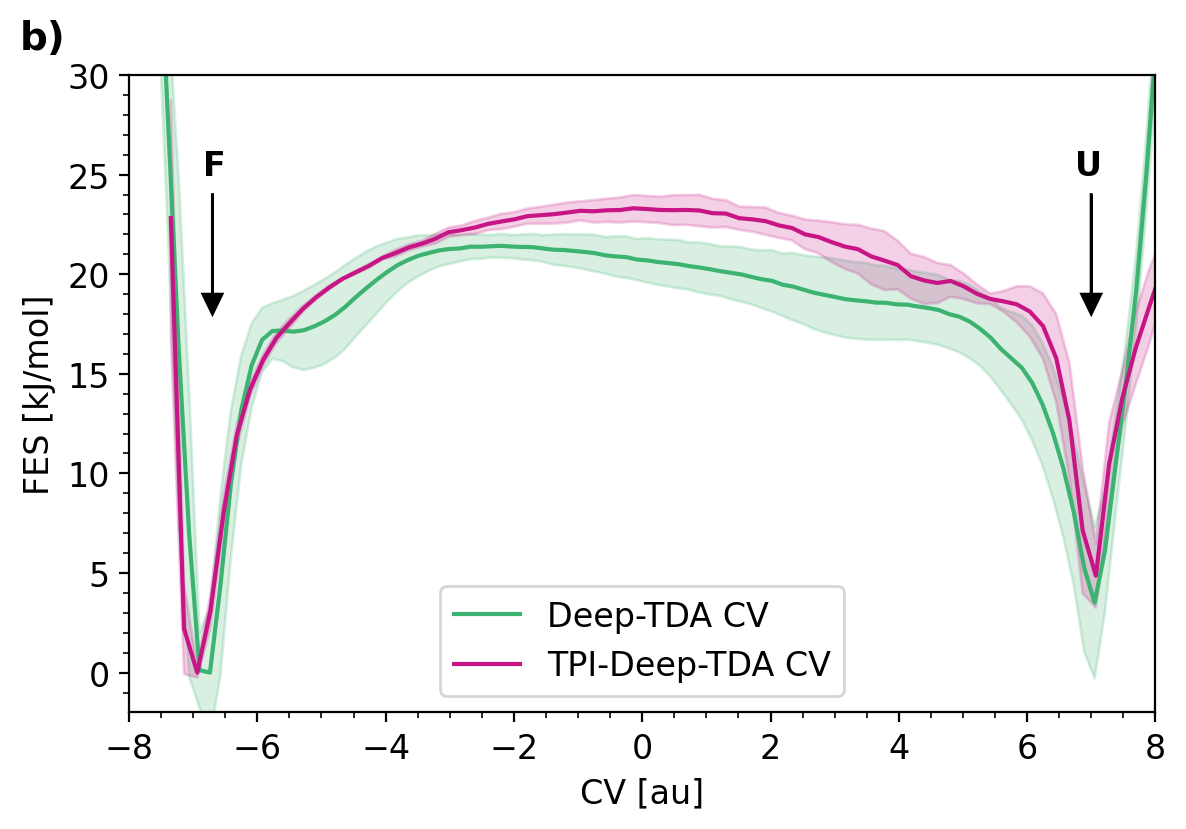}
    \end{minipage}
    \caption{\textbf{(a)} Comparison of the convergence with simulation time of the free energy of Chignolin folding as obtained from OPES simulation with Deep-TDA and TPI-Deep-TDA CVs. The solid line and the shaded region report respectively the standard deviation computed from three independent trajectories. The dashed line reports the reference value obtained with long unbiased simulations\cite{lindorff2011fast}. The  $\pm$0.5k$_B$T range around this value is marked by the thin dotted lines.
    \textbf{(b)} 1D projection of the free energy surface computed biasing the standard Deep-TDA and the TPI-Deep-TDA CVs. As the two CVs are trained to map the folded (F) and unfolded (U) states to the same positions, each 1D-FES is directly projected along the corresponding CV.}
    \label{fig:chignolin_results}
\end{figure*}

The TPI-Deep-TDA CV, trained on $C_{\alpha}$ contacts, results in a better convergence with a tighter confidence interval as compared to the standard Deep-TDA (Fig. \ref{fig:chignolin_results}). In less than 200 ns, the free energy difference between the folded and the unfolded state converged within one $k_BT$ of the results obtained from $\sim$ 100 $\mu$s long unbiased simulation \cite{lindorff2011fast}. In contrast, using standard Deep-TDA CV, it took around 500 ns to reach convergence. Moreover, the uncertainty in the free energy difference between the folded and the unfolded states was larger than that of TPI-Deep-TDA.
(See Supporting Information (SI) Fig. S2). One of the reasons for the increased efficiency is due to the fact that when using TPI-Deep-TDA, the exploration of the FES is limited to the minimum free energy path as observed in the 2D model potential. 

A reflection of the superior quality of our CV is the height of the barrier of the free energy projected on the TPI-Deep-TDA being higher than the standard Deep-TDA projection\cite{bal2020free} as shown in panel b of Fig.\ref{fig:chignolin_results}. 

In addition to the folding free energy, the Chignolin unfolding time has was also obtained from the 20 unbiased folding events sampled from the OPES-Flooding simulations. As reported in Table \ref{tbl:kinetics} the results are in agreement with those obtained with long unbiased simulations\cite{lindorff2011fast}. The computational efficiency of the kinetics calculation, measured in terms of the acceleration factor, is almost 2 times than that of the Deep-TICA and Deep-LDA CVs used in our earlier work \cite{ray2022rare}. This result highlights the superiority of the Deep-TDA CV for computing rates using OPES-Flooding simulations.

\begin{table}[h!]
\resizebox{\textwidth}{!}{\begin{tabular}{|l|l|l|l|l|l|l|}
\hline
System                                                                       & \begin{tabular}[c]{@{}l@{}}Number \\ of runs\end{tabular} & \begin{tabular}[c]{@{}l@{}}Mean first \\passage time ($\tau$) \end{tabular}      & $p$-value \textsuperscript{\emph{a}} & \begin{tabular}[c]{@{}l@{}}95\% Confidence \\ interval\textsuperscript{\emph{b}}\end{tabular} & \begin{tabular}[c]{@{}l@{}}Acceleration\\ factor\end{tabular} & \begin{tabular}[c]{@{}l@{}}Reference \\ value\end{tabular} \\ \hline
\begin{tabular}[c]{@{}l@{}}Chignolin \\ (unfolding)\end{tabular}             & 20                                                        & 3.08 $\mu$s & 0.665     & 1.94 - 4.70 $\mu$s                                                        & 316                                                           & 2.2 $\pm$ 0.4 $\mu$s (Ref. \citen{lindorff2011fast})                                      \\ \hline
\begin{tabular}[c]{@{}l@{}}Host-Guest \\ unbinding \\ (OAMe-G2)\end{tabular} & 13                                                        & 3.78 ms     & 0.338     & 1.82 - 5.52 ms                                                        & 1.5$\times 10^6$                                              & 2.02 ms  (Ref. \citen{debnath2022computing})                                                  \\ \hline
\end{tabular}}

\textsuperscript{\emph{a}}  The $p$-values are computed from 2 stample Kolmogorov-Smirnoff test \cite{salvalaglio2014assessing}. \\
\textsuperscript{\emph{b}} The 95\% confidence intervals were computed as $\lbrace \frac{2\sum_i^n t_i}{\chi_{2n}^2 (0.975)}, \frac{2\sum_i^n t_i}{\chi_{2n}^2 (0.025)} \rbrace$ as suggested by Kaminsky \cite{kaminsky1972confidence}. The $t_i$ refers to the rescaled time for the $i$-th transition, and $\chi_{2n}^2 (\alpha)$ is critical value of two-tailed $chi$-squared test with $2n$ degrees of freedom and $p$ = $\alpha$.

\caption{The kinetics obtained using OPES-Flooding during the training of the TPI-Deep-TDA CV. }
\label{tbl:kinetics}
\end{table}

\subsection{Ligand receptor binding}
\begin{figure*}[h!]
    \begin{minipage}{0.49\linewidth}
        \includegraphics[width=\linewidth]{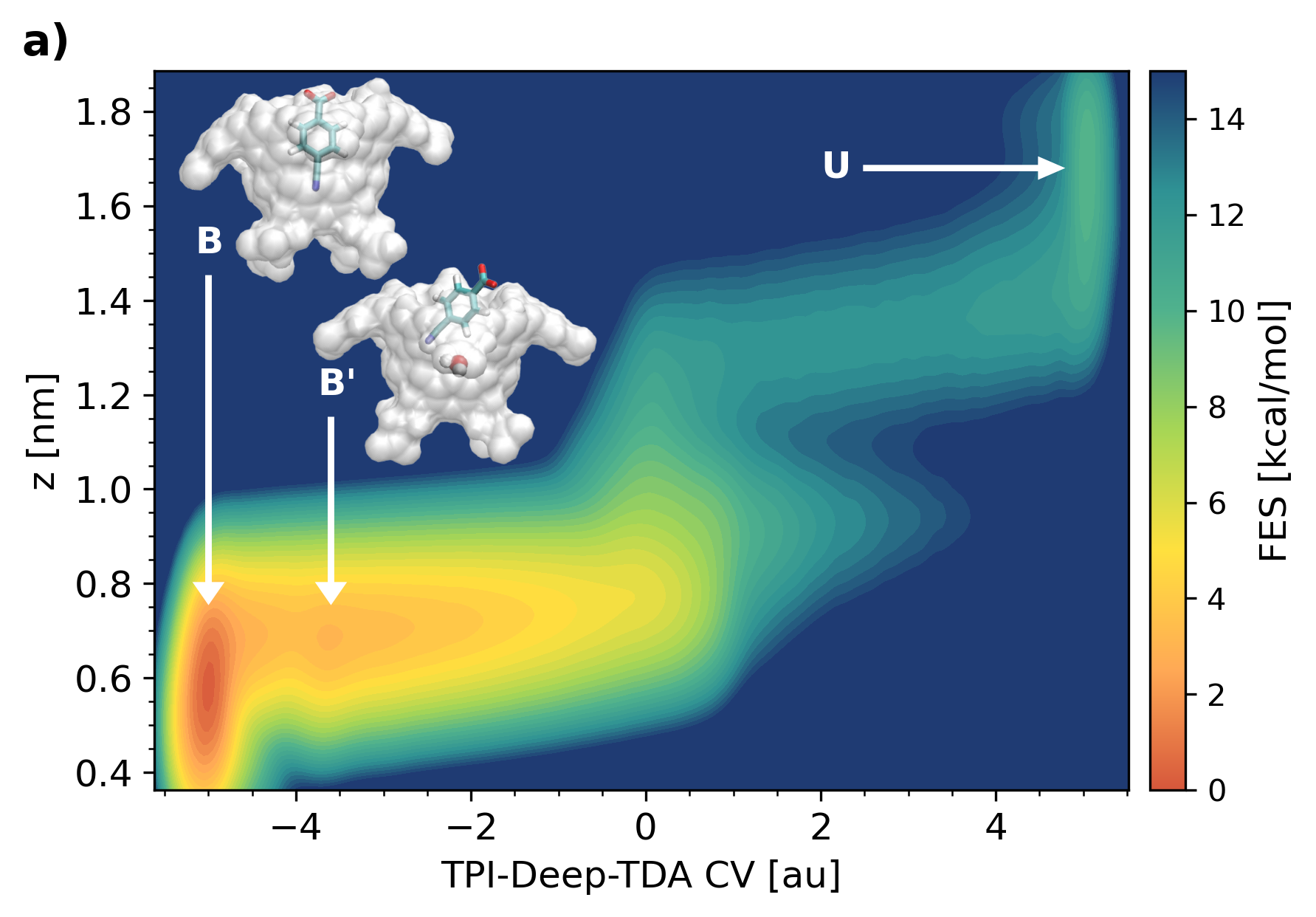}
    \end{minipage}
    \hspace*{\fill}
    \begin{minipage}{0.49\linewidth}
        \includegraphics[width=\linewidth]{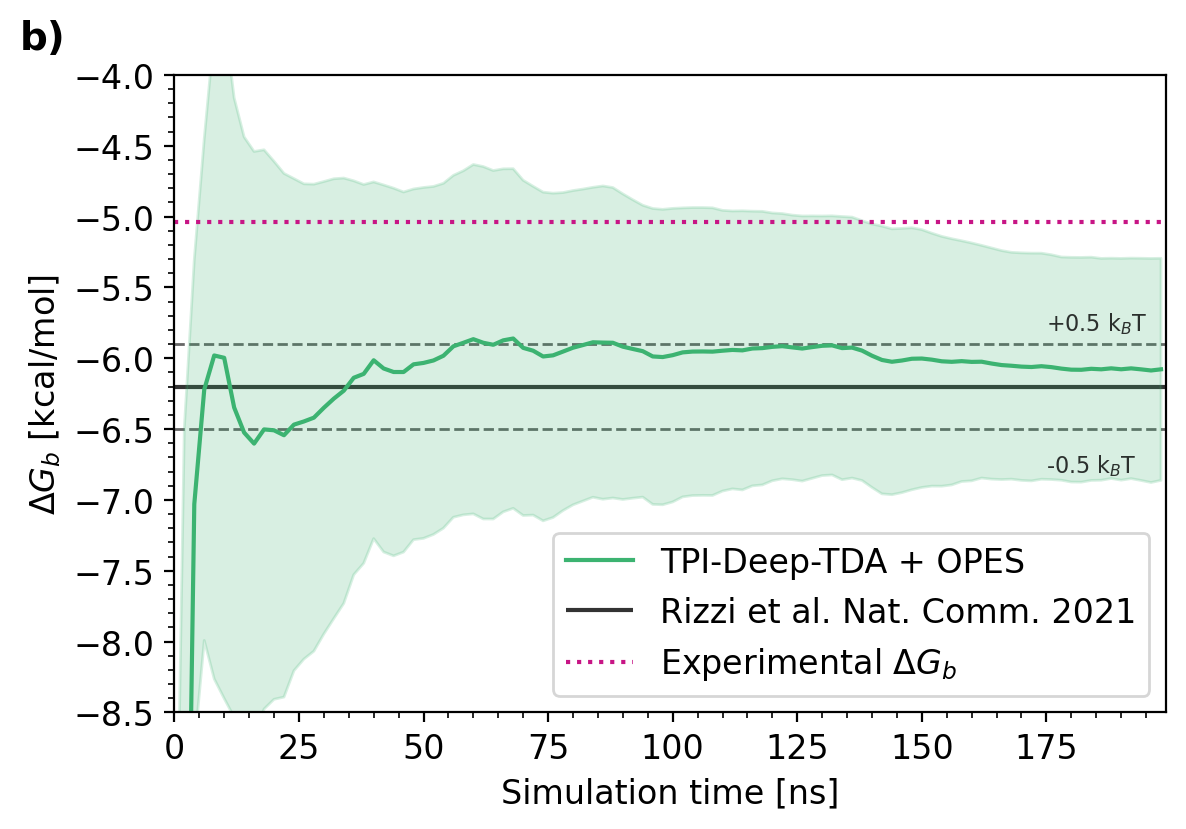}
    \end{minipage}
    \caption{\textbf{(a)} 2D free energy surface projected along the TPI-Deep-TDA CV and the vertical component z of the distance between the ligand and the binding site. Three metastable states are visible: unbound (U), bound (B) and semi-bound (B'). Representative structures of the B and B' states are provided in the inset. 
    \textbf{(b)} Convergence with time of binding free energy of G2 guest in OAMe octa-acid host using TPI-Deep-TDA CV. The solid line and the shaded region report respectively the standard deviation computed from three independent trajectories. As a reference, we report in black the computational value obtained in Ref.\cite{rizzi2021role} with a similar setup and in pink the experimental value from Ref.\cite{yin2017overview}. The dashed lines give the $\pm$0.5k$_B$T range on the computational reference value.}
    \label{fig:sampl5_2d_fes}
\end{figure*}

Lastly, we tested our TPI-Deep-TDA approach for the binding of G2 guest to the OAMe octa-acid host, used in the SAMPL5 challenge. Following our approach, the binding free energy converges rapidly (in less than 50 ns) to a value ($-6.08 \pm 0.78$ kcal/mol) similar to the result obtained by Rizzi et al. ($-6.19 \pm 0.08$ kcal/mol)\cite{rizzi2021role} (Fig. \ref{fig:sampl5_2d_fes}, panel b) who studied this system in great detail using identical force field parameters. The computed binding free energy is not in perfect agreement with the experimental results (-5.04 kcal/mol), likely due to the approximate nature of the empirical force field. However, we could exactly reproduce the result of Rizzi et al. \cite{rizzi2021role}, confirming the usefulness of our CV and we also observed frequent transitions between the bound and the unbound states during the course of the OPES simulation (see SI Fig. S7). 

In panel a of Fig. \ref{fig:sampl5_2d_fes}, we show the average free energy surface from three independent simulations, projected along the ligand-receptor distance ($z$) and the TPI-Deep-TDA CV. 
As in Ref. \citen{rizzi2021role}, we found that there is a bound state (B) and an intermediate semi-bound configuration (B'). This latter state appears as a shallow minimum close to the B state.  In the B' state, the binding pocket is occupied by one water molecule, which prevents the guest to attain the minimum energy-bound configuration (Fig. \ref{fig:sampl5_2d_fes}, inset of panel a). The nature of this state is discussed in detail in previous work\cite{rizzi2021role}. 
The free energy difference between this shallower minimum and the true bound state was reported to be approximately 2 kcal/mol, which is in agreement with our results.

In addition to the binding free energy,  we computed the ligand residence time using the 13 unbinding events observed in our OPES-Flooding simulations. The result is in excellent agreement with the work of Debnath and Parrinello \cite{debnath2022computing} (Table \ref{tbl:kinetics}) that used the Gaussian Mixture based enhanced sampling (GAMBES) scheme \cite{debnath2020gaussian}. Our ability to obtain millisecond timescale ligand residence time from nanosecond long simulations (i.e acceleration factor $\approx 10^6$) bodes well for future applications on ligand-binding problems of practical interest. 

\section{Conclusions}
In this work, we propose an improvement to the machine learning based collective variable discovery procedure by incorporating information about the transition path ensemble. We discuss a two-stage protocol where, first, an approximate CV trained on metstable state information is used to generate transition paths, these pathways are then utilized to develop a more accurate CV that can distinguish the initial state, the final state, and the transition state. Our approach was tested on the barrier crossing in a 2D model potential, the folding and unfolding of chignolin miniprotein, and the binding of a small molecule ligand to a synthetic host. In this diverse set of problems, the CV designed using our novel TPI-Deep-TDA protocol resulted in a highly accurate prediction of the free energy difference between the metastable states as well as led to a quicker convergence of results in comparison to the standard Deep-TDA CV trained only on the metastable states. The faster convergence of the free energy surface compensates for the additional computing effort invested in sampling the transition path ensemble for the initial training. However, this effort brings in the additional benefit of being able to compute kinetic rates as a part of the CV designing procedure. This is possible as an accurate description of transition state region is not essential for the recovery of accurate kinetics using infrequent metadynamics \cite{tiwary2013metadynamics} or OPES-Flooding \cite{ray2022rare}.
We demonstrated that biasing along a transition path informed CV can direct the flux of transitions through the minimum free energy path, leading to a more accurate description of the transition pathways, and facilitating the understanding of the atomistic mechanisms of complex processes.


Considering the multitude of benefits of including the transition path information in collective variables discovery, and the relatively simple and semi-automated training procedure, TPI-Deep-TDA CV may find a wide range of applications in studying rare-events processes in biology, materials science, and chemistry.


\begin{acknowledgement}
The authors thank Luigi Bonati and Valerio Rizzi for stimulating discussions, and Narjes Ansari for sharing input files of the host-guest system from Ref. \citen{rizzi2021role}. The authors thank D.E. Shaw Research for sharing the input files and trajectories of Chignolin from Ref. \citenum{lindorff2011fast}. The authors declare no competing financial interest.

\end{acknowledgement}

\section{Data Availability Statement}
The input files for all simulations performed in this work and sample codes for training the TPI-Deep-TDA CV are provided in the GitHub repository: \url{https://github.com/dhimanray/TPI_deepTDA.git}. The input files will also be made available through the PLUMED NEST repository\cite{plumed2019promoting}.



\begin{suppinfo}
Computational details, neural network training protocol, supplementary results, and Figure S1-S8 are available in the supporting information (SI).

\noindent

\end{suppinfo}

\bibliography{bibliography}

\end{document}